\documentclass{optica-article}

\journal{opticajournal} 

\articletype{Research Article}

\usepackage{lineno}

\begin{document}

\title{Compact and robust optical frequency reference module based on reproducible and redistributable optical design}

\author{Jiwon Wi,\authormark{1,2} Taehee Kim,\authormark{1} and Junki Kim\authormark{1,2,3,*}}

\address{\authormark{1}SKKU Advanced Institute of Nanotechnology (SAINT) \& Department of Nano Science and Technology, Sungkyunkwan University, Suwon, 16419, Korea\\
\authormark{2}Department of Nano Engineering, Sungkyunkwan University, Suwon, 16419, Korea\\
\authormark{3}Department of Quantum Information Engineering, Sungkyunkwan University, Suwon, 16419, Korea}

\email{\authormark{*}junki.kim.q@skku.edu} 

\begin{abstract*} 

Stabilized optical frequency references (OFRs) are indispensable for atom-based quantum technologies, optical communications, and precision metrology.
As these systems become more sophisticated, demands for compactness, robustness, and straightforward reproduction have grown.
In this work, we present a robust 19-inch rack-mountable OFR module designed via a web-based CAD workflow that allows straightforward redistribution and reproduction.
Its optical subsystem, designed based on a modeled laser beam path, places optical elements with sub‑millimeter accuracy on a custom‑machined aluminum plate, allowing straightforward assembly without extensive alignment and providing high mechanical stability. 
The module maintains frequency‑stable operation for several months without user intervention and exhibits high robustness to mechanical vibrations up to 4g.
All design files, including mechanical and optical metadata, are openly shared for straightforward reproduction and adaptation.

\end{abstract*}

\section{Introduction}

Stabilized optical frequency references (OFR)\cite{ye_sub-doppler_nodate,holzwarth_optical_2000,hollberg_optical_2001,bartels_femtosecond-laser-based_2005,gill_optical_2005,hall_nobel_2006} are essential components in many atomic molecular optics (AMO) experiments, including quantum technologies based on ions\cite{leibfried_quantum_2003, bruzewicz_trapped-ion_2019} or neutral atoms\cite{henriet_quantum_2020, wintersperger_neutral_2023}, optical communications\cite{couteau_applications_2023, cozzolino_high-dimensional_2019} and metrology\cite{pezze_quantum_2018, giovannetti_advances_2011}.
Moreover, recent advances in optical technology have extended the application of precision optical devices to harsh environments, including field operation\cite{zhang_microrod_2019,davila-rodriguez_compact_2017} and even outer space\cite{schuldt_high-performance_2016,schkolnik_jokarus_2017,strangfeld_prototype_2021,dinkelaker_autonomous_2017}.
Such an environment requires optical systems that are both portable \cite{pogorelov_compact_2021,zhang_compact_2018,burke_compact_2005,strangfeld_compact_2022} and robust against external noise\cite{pahl_compact_2019,chen_compact_2014}.

Conventional OFR assemblies often rely on custom optical layouts and manual alignment, which limits reproducibility and complicates redistribution across different laboratories.
As these systems are increasingly deployed in broad applications, a design approach that supports straightforward reproduction and adaptation is becoming essential\cite{myers_qubit_2025}. 
Such an approach facilitates collaborative development, simplifies maintenance, and enables consistent performance across diverse environments.

In this work, we present a robust OFR module with a design that enables straightforward reproduction and redistribution.
Reproducibility is achieved through a standardized optical layout in which all components are positioned according to a predefined beam path on a custom‑machined aluminum plate, enabling assembly with sub‑millimeter precision and minimal alignment.
The complete design, including component relationships and a simplified graphical representation, is available through a web-based CAD platform and a public repository.
The entire module is assembled in a standard 19‑inch rack‑mountable form factor, providing enhanced mechanical stability and supporting fully remote operation.
We successfully built two identical modules with the same design, each operating stably for several months without user intervention.

\section{Module design}
\begin{figure}
    \centering
    \includegraphics[width=0.7\textwidth]{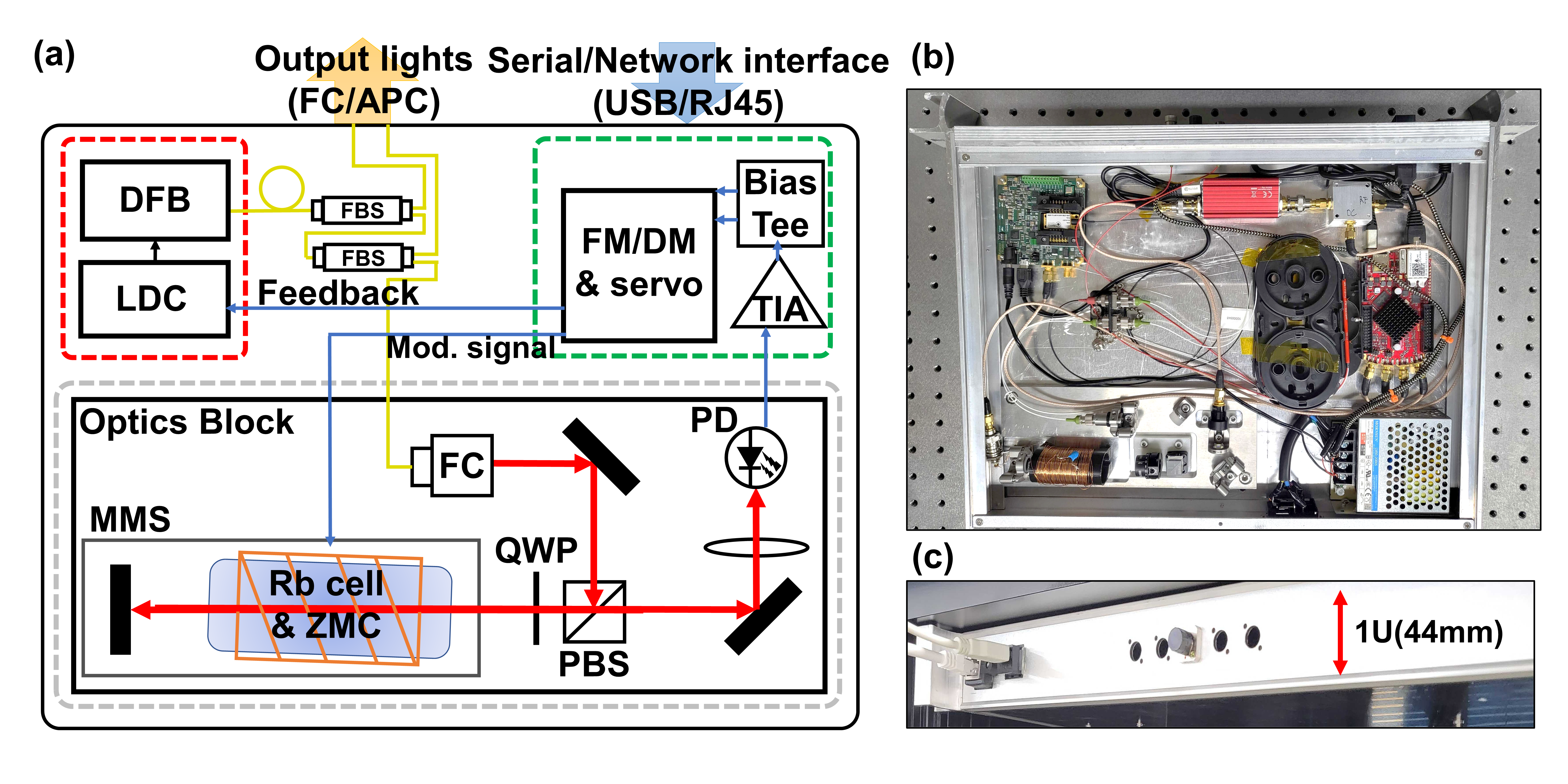}
    \caption{(a) Schematic of the OFR module. The laser source and controller (red dashed box), spectroscopic optics (gray dashed box), and control electronics (orange  dashed box) are integrated into a single rack-mountable case. The electronics and laser controller are powered by an internal 5 V DC power supply (not depicted) and remotely controlled via TCP/IP and serial communications. DFB, distributed-feedback laser; LDC, laser diode controller; FBS, fiber beam splitter; FC, fiber collimator; PBS, polarizing beam splitter; QWP, quarter- wave plate; PD, photodiode; ZMC, Zeeman modulation coil; MMS: mu-metal shield; TIA, transimpedance amplifier; FM/DM, frequency modulation/demodulation. (b,c) Photographs of the constructed OFR module. All the optical and electrical components are commercially available off-the-shelf parts.}
    \label{fig:overview}
\end{figure}

The OFR module design incorporates all the necessary optical and electronic components for a frequency-stabilized laser source, referenced by the rubidium D2 transition\cite{terra_ultra-stable_2016,ye_hyperfine_1996,imanishi_frequency_2005}.
The module comprised three main subsystems; laser source, optical submodule for spectroscopy, and control electronics, as shown in Fig.\ref{fig:overview}.
The module operates standalone by plugged to AC power and can be remotely controlled to configure settings or lock the laser.

To accomplish robustness against the external environment and compactness of the module, we integrated all the components into a compact 1U standard 19" rack unit. 
The whole module design is based on online platform, not only to help reproduction of the module but also to assist redistribution and version control.
The optical and electronic components in the module are readily available commercial off-the-shelf (COTS) parts which enhances the reproducibility.
The entire bill of materials (BOM) and design files of the module are uploaded in the public repository \cite{noauthor_queti-at-skkuofr-module_nodate} (\href{https://github.com/queti-at-skku/ofr-module}{https://github.com/queti-at-skku/ofr-module}).

\subsection{Optical subpart and reproducible blueprint} 

\begin{figure}
    \centering
    \includegraphics[width=0.9\textwidth]{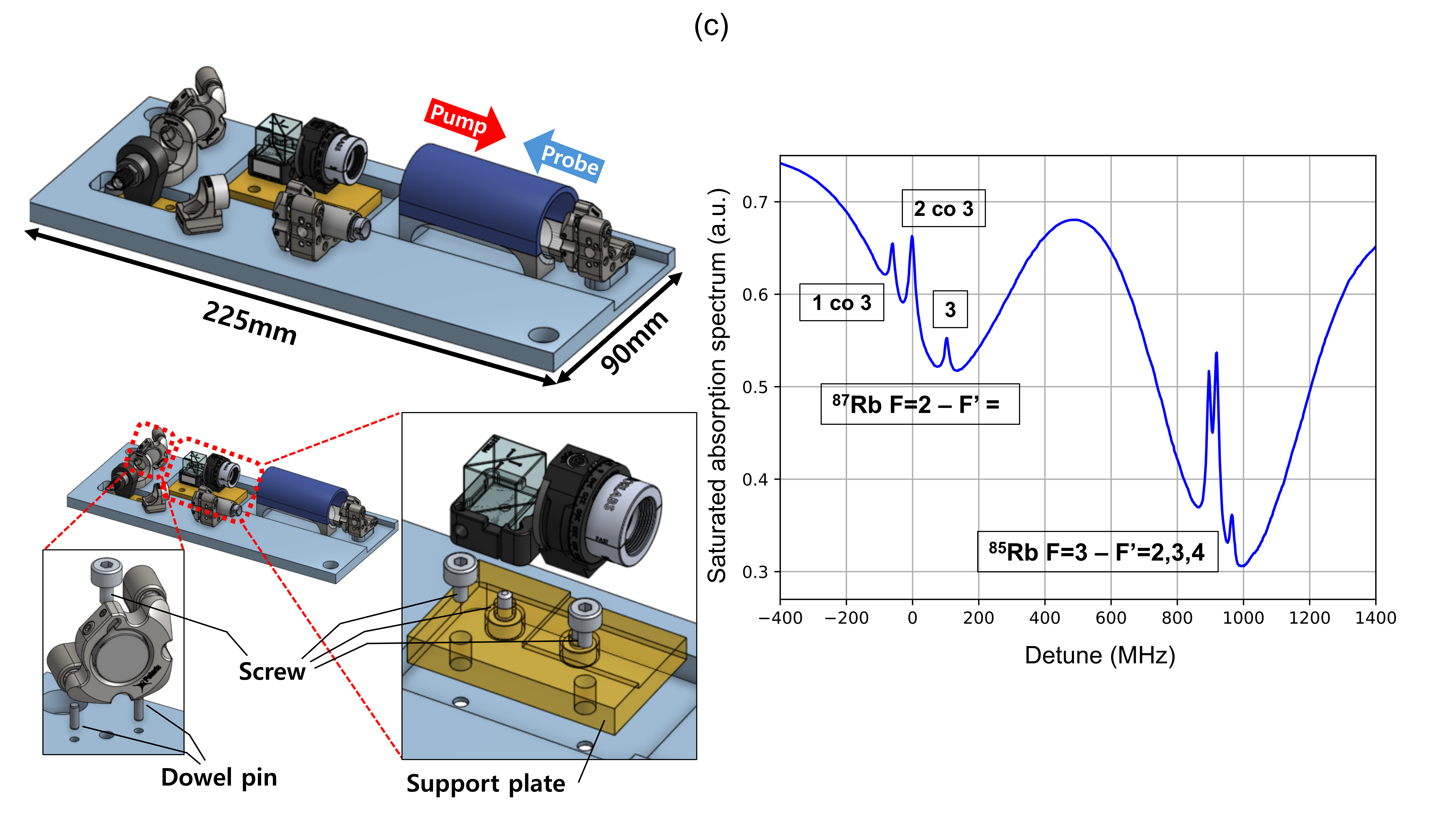}
    \caption{
    (a) Overall design of the SAS optical block. All optical components for SAS are mounted on an compact 225mm $\times$ 90mm $\times$ 8mm aluminum plate. Pump and probe beam are counterpropagated through Rb vapor cell to generate a saturated absorption spectrum. (b) Optical components are either directly mounted on the base plate or secured via custom 3D‑printed adapters. Precise positioning is achieved using additional dowel pins, which ensure reproducible alignment. (c) Saturated absorption spectrum of rubidium vapor cell. The curve shows the Doppler-free and crossover resonance peaks of rubidium 85 and 87. The OFR module locks the 780 nm laser frequency to the $F' = 2~ \textrm{co}~ 3$ crossover resonance of rubidium 87 (middle peak) at 384.227981THz.}
    \label{fig:conceptual}
\end{figure}

Figure \ref{fig:conceptual} (a) shows the 3D rendering of the optical subpart for spectroscopy.
The optical block serves as the core of the module, producing the spectroscopic error signal referenced by the rubidium atoms in the vapor cell.
The module employs saturated absorption spectroscopy (SAS) \cite{preston_dopplerfree_1996} configuration, where the weak probe beam measures the frequency-dependent attenuation through the rubidium vapor cell while the strong counter-propagating pump beam influences the state population of the atoms.
The vapor cell orientation is deliberately rotated by a small angle ($\approx 2 ^{\circ}$) to prevent back-propagation of the probe beam into the laser source. 
Finally, the probe beam is focused by a lens and collected by the photodetector and the intensity of the transmitted probe beam can reveal the Doppler-free spectroscopy of rubidium atoms.
Figure \ref{fig:conceptual} (c) shows the resolved SAS spectrum of rubidium-85 and rubidium-87 D2 lines, labeled by hyperfine $F$ number of the excited state.

The entire optical subsystem was designed using CAD to ensure reproducibility and ease of redistribution.
In the design, the laser beam passes the center of each optical element, and the mirror angles are aligned to ensure the proper beam reflection.
To meet the design objectives, the optical elements in the free-space optics must be positioned with sub-millimeter precision. 
We machined the base plate from aluminum alloy and incorporated dowel pin holes to precisely fix the positions of the optical components (Fig. \ref{fig:conceptual} (b)).
For elements that cannot be mounted directly onto the base plate, such as the vapor cell and polarizing beam splitters, we used custom 3D-printed adapter parts made of polylactic acid (PLA) material.
All machined and printed parts are reproducible via STEP or STL format and the remaining optical elements are standard COTS components, which can be easily replaced.
As a result, the production and assembly of the optical subpart are straightforward and reproducible.
Using this reproducible approach, we successfully built two additional modules with identical design and performance.

\begin{figure}
    \centering
    \includegraphics[width=0.7\textwidth]{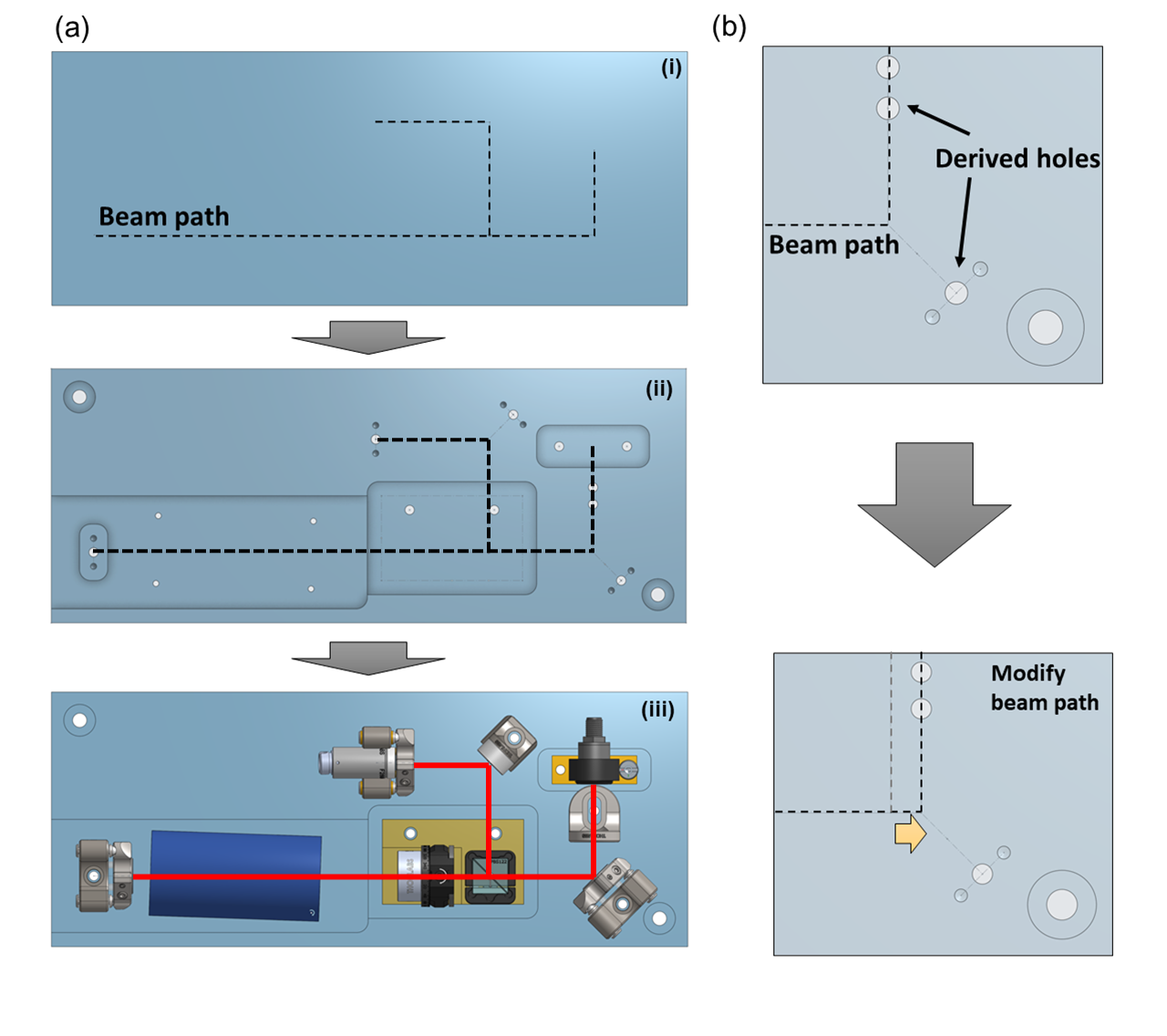}
    \caption{ (a) Design step of the optical block. First, the beam path is defined according to the desired functionality (i). Based on this path, the positions of the optical elements and the corresponding mounting holes and cutouts are derived (ii). The resulting base plate can host all optical components with high precision, requiring minimal alignment adjustment during assembly (iii). 
    (b) When design modifications are needed, simply adjusting the beam path automatically updates all derived features. Iterating this process enables optimization of the layout and reduction of mechanical conflicts.}
    \label{fig:cad}
\end{figure}

The core of this design process lies in defining the optimal beam path for the optical subsystem. 
While the beam path in actual optical systems is determined by the optical elements, in the design phase, it could be more effective to first establish the desired beam path, from which the position and orientation of optical elements can subsequently be derived.
Optimizing the beam path is, in essence, equivalent to optimizing the entire optical design.
Figure \ref{fig:cad} (a) illustrates the detailed steps of our design process. 
Once the beam path is defined, the required optical elements can be positioned accordingly to align with its trajectory.
Since the number and placement of components depend on the beam path, the design can be iteratively refined to avoid mechanical conflicts and eliminate design redundancies.
After optimization, Our spectroscopy submodule consists of just nine optical elements, with dimensions of 225 mm by 90 mm.

Our design process is implemented on 'Onshape'\cite{business_onshape_nodate}, a web‑based CAD platform that supports version control and collaborative designs.
In this framework, the beam path is modeled as line segments representing the intended laser trajectory. 
Optical components are then placed along this path, with their positions and orientations derived by CAD constraints to the line elements.
From these constraints, the necessary mounting features, including tapped holes, dowel pins, and plate cutouts, are automatically derived according to the specifications of each component. 
Because all positional data originates from the beam path, any design modification can be implemented simply by adjusting the path line elements, facilitating both iterative refinement and design optimization (Fig \ref{fig:cad} (b)).

\begin{figure}
    \centering
    \includegraphics[width=0.7\textwidth]{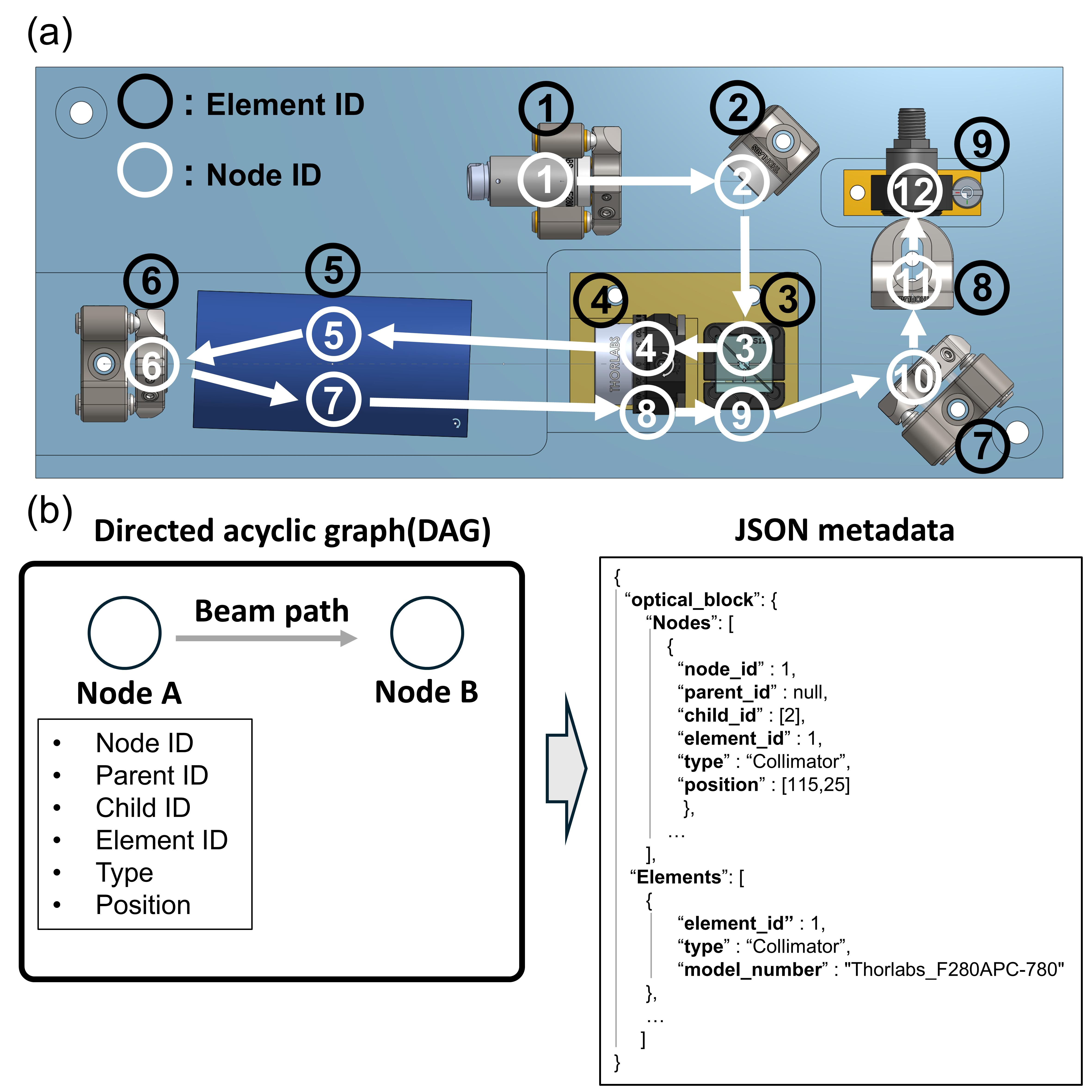}
    \caption{
    (a) Representation of the optical block using graph mapping. 
    The beam path is described as a sequential flow of nodes (white), with each optical component identified by a corresponding element ID (black).
    The directed edges indicate the beam propagation through the optical system. 
    (b) Conversion of the optical layout into a directed acyclic graph (DAG) and corresponding JSON metadata.
    Each node stores attributes such as parent/child relationships, element ID, component type, and spatial position, enabling systematic design description and reproducibility.}
    \label{fig:optical_DAG}
\end{figure}

The full description of the beam path can be effectively captured as graph form, specifically as a directed acyclic graph (DAG).
A DAG is a type of graph where edges have a direction, and no cycles are present, ensuring that each path follows a single direction without returning to any prior node.
This structure is particularly suitable for modeling optical beam paths at an abstract level, as the directed edges align naturally with the light’s inherent directionality, while the explicit ordering within a DAG allows for a clear understanding of the design intent behind the beam path. 
Additionally, this approach enhances comprehension of the dependencies among optical components, aiding in both design and troubleshooting.
In the graph, each node corresponds to an optical component, ranging from lenses to mirrors and other elements, while each directed edge indicates the path of light between elements, providing a systematic representation of the optical system (Fig. \ref{fig:optical_DAG}).
The graph metadata can be compiled in JSON format, and by distributing the structural metadata, the connectivity between elements remains clearly defined, even with more sophisticated systems. 

\subsection{Laser source and control electronics}
We used a single frequency distributed feedback laser diode (Toptica EYP-DFB-0780-00010-1500-BFY12-0005) controlled by a digital butterfly laser diode controller (Coheron CTL200) as the light source. 
The laser diode was coupled to a polarization-maintaining optical fiber and interfaced with a standard FC/APC fiber connector, which was butt-coupled to fused fiber splitters (Thorlabs PN780R5A1) and routed to the spectroscopy setup.
We modularized the SAS setup in a small aluminum plate (225 mm × 90 mm). All the optics on the plate were fitted to a half-inch scale and fixed by an M4 screw and 2.1 mm dowel pin. 
A rubidium vapor cell was mounted in a 3D-printed structure surrounded by a solenoid. 
The solenoid received radiofrequency signals from control electronics and generated a magnetic field.

The laser passed through the optical block and was detected by a photodiode (Thorlabs SM05PD3A). 
The spectroscopy signal from the photodiode was amplified by a transimpedance amplifier. The signal was split into two parts by using a bias tee, obtaining a DC spectroscopy signal for monitoring and AC spectroscopy signal for frequency locking. 
The two signals were directed to a compact FPGA-based digital electronic system (Redpitaya STEMlab 125-14), which performed signal monitoring, modulation, demodulation, and servo for proportional–integral–derivative (PID) control feedback. 
The signal configuration and PID control setup in the feedback loop was easily controlled via the Linien open-source laser locking software\cite{wiegand_linien_2022}.

To lock the laser frequency at the peaks of the rubidium-saturated absorption curve, a zero-crossing error signal was indispensable.
By frequency-modulation spectroscopy\cite{bjorklund_frequency-modulation_1980,nakanishi_frequency-modulation_1987,zi_laser_2017}   
, the module generated an error signal using Zeeman sub-level modulation [16] for frequency stabilization. 
To generate the error signal, a Zeeman sub-level modulation can be used\cite{corwin_frequency-stabilized_1998}. 
The modulation coil was designed to have a resonance frequency of 250 KHz. The Zeeman shift at the coil was 1 MHz. After modulating the atomic transition, the electronic system demodulated the signal and generated the error signal of the saturated spectrum. 
By applying Zeeman modulation instead of direct modulation at the laser diode, we removed the unwanted modulation frequency in the laser.

\section{System characteristics \& performance} 
\subsection{System characteristics}
The complete system is packaged in a standard 437 mm × 280 mm rack-mountable enclosure with a height of 1U (44.4 mm), corresponding to the lowest standard rack unit. 
The module weighs 3.98 kg, allowing transport by a single person. 
The frequency-stabilized laser is delivered through two output ports, providing optical powers up to 40 uW and 1 mW, respectively.

\begin{table}[htbp]
\centering
\begin{tabular}{cc}
\hline
\textbf{Specification} & \textbf{Value} \\
\hline
Size (width × depth) & 437 mm × 280 mm \\
Height & 1U (44.4 mm) \\
Weight & 3.98 kg \\
Power consumption & 7 W \\
\hline
\end{tabular}
\caption{Specifications of the OFR module. U denotes the standard rack unit (44.4 mm). The module is compatible with a standard 19'' rack.}
\end{table}

The complete system can be controlled by an external control computer. 
Two communication ports (USB and RJ45) are installed on the front panel.
Serial communication allows configuration of the laser diode, including laser diode current and temperature control. 
Through a TCP/IP interface, the user can operate the electronics via the Linien graphic interface. 
The parameters for laser locking, such as frequency scan range, modulation frequency, and PID settings, can be also adjusted, and the system status can be monitored remotely.

\subsection{Evaluation tests}
We performed laser frequency tracking using a wavelength meter (WLM; HighFinesse WS8-10).
After the laser is locked to the target frequency, we recorded the time-series frequency data using the WLM for 30 min.
The long-term stability of the reference laser was evaluated by calculating the Allan deviation\cite{allan_statistics_1966,chen_stable_2022}, defined as 

\begin{equation}
    \sigma^2(\tau) = \frac{1}{2(M-2n+1)}\sum_{j=0}^{M-2n}(\Bar{y}_{j+n}-\Bar{y_j})^2
\end{equation}

\noindent where $\Bar{y}_{j}$ is normalized frequency value, and $M$ is number of total data points, $\Delta$ is time interval for measurement(200ms in this study), and average time $\tau$ is defined as $\tau = n\Delta$.

\begin{figure}
    \centering
    \includegraphics[width=\textwidth]{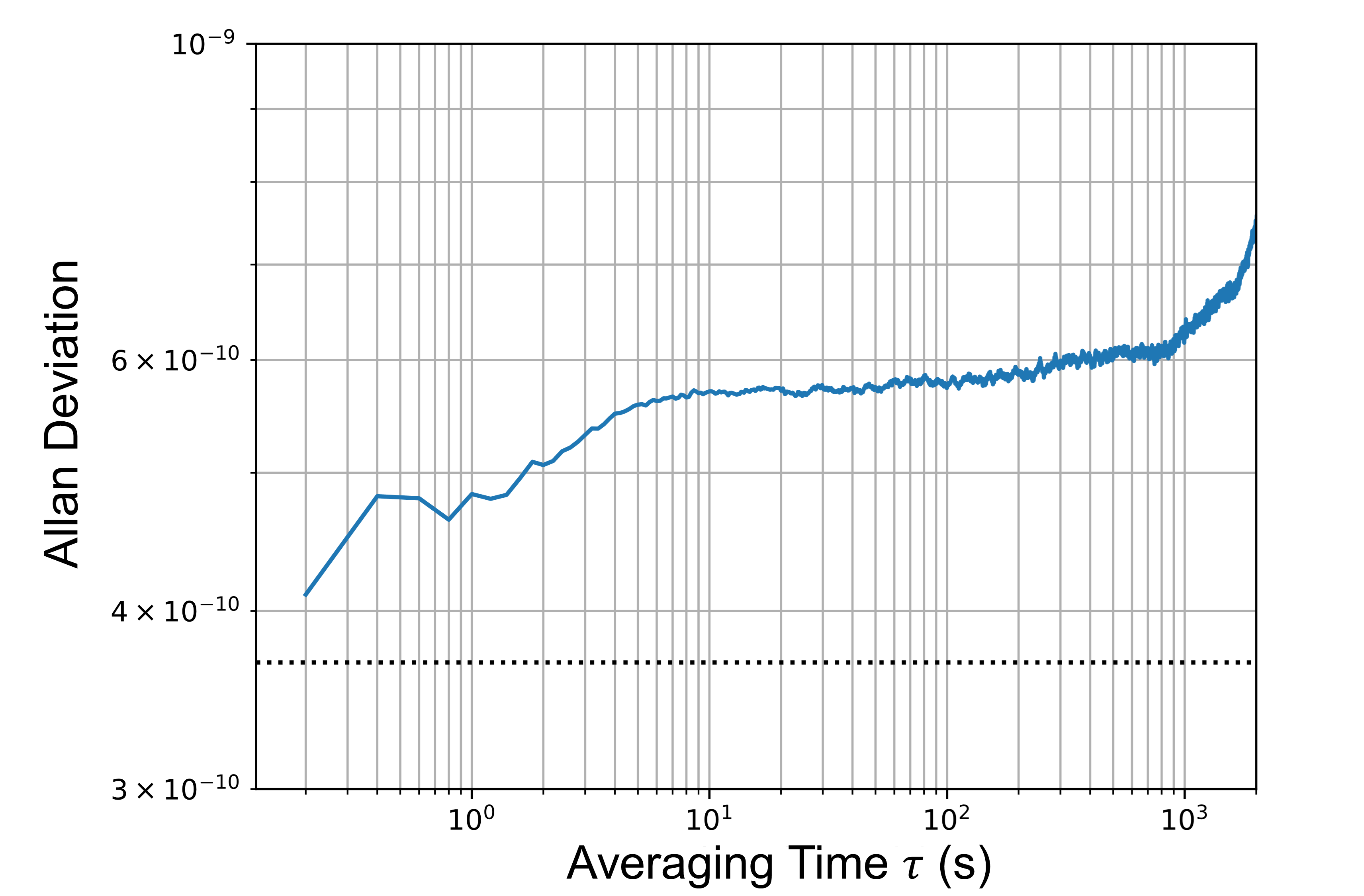}
    \caption{Allan deviation of frequency-stabilized laser measured by WLM. The dotted line represents the measurable lower bound imposed by the WLM resolution (400 kHz). The drift after $10^3$ s is caused by a device calibration shift}
    \label{fig:allan}
\end{figure}
 
Figure \ref{fig:allan} shows the Allan deviation of the reference laser.
The lower bound imposed by the WLM resolution was calculated to be $3.68 \times 10^{-10}$ at 780 nm. 
The results confirm that the module maintains stable laser frequency within this limit.
The increase observed beyond $10^3$ s originates from calibration drift of the wavelength meter rather than intrinsic instability of the system.
 
\begin{figure}
    \centering
    \includegraphics[width=\textwidth]{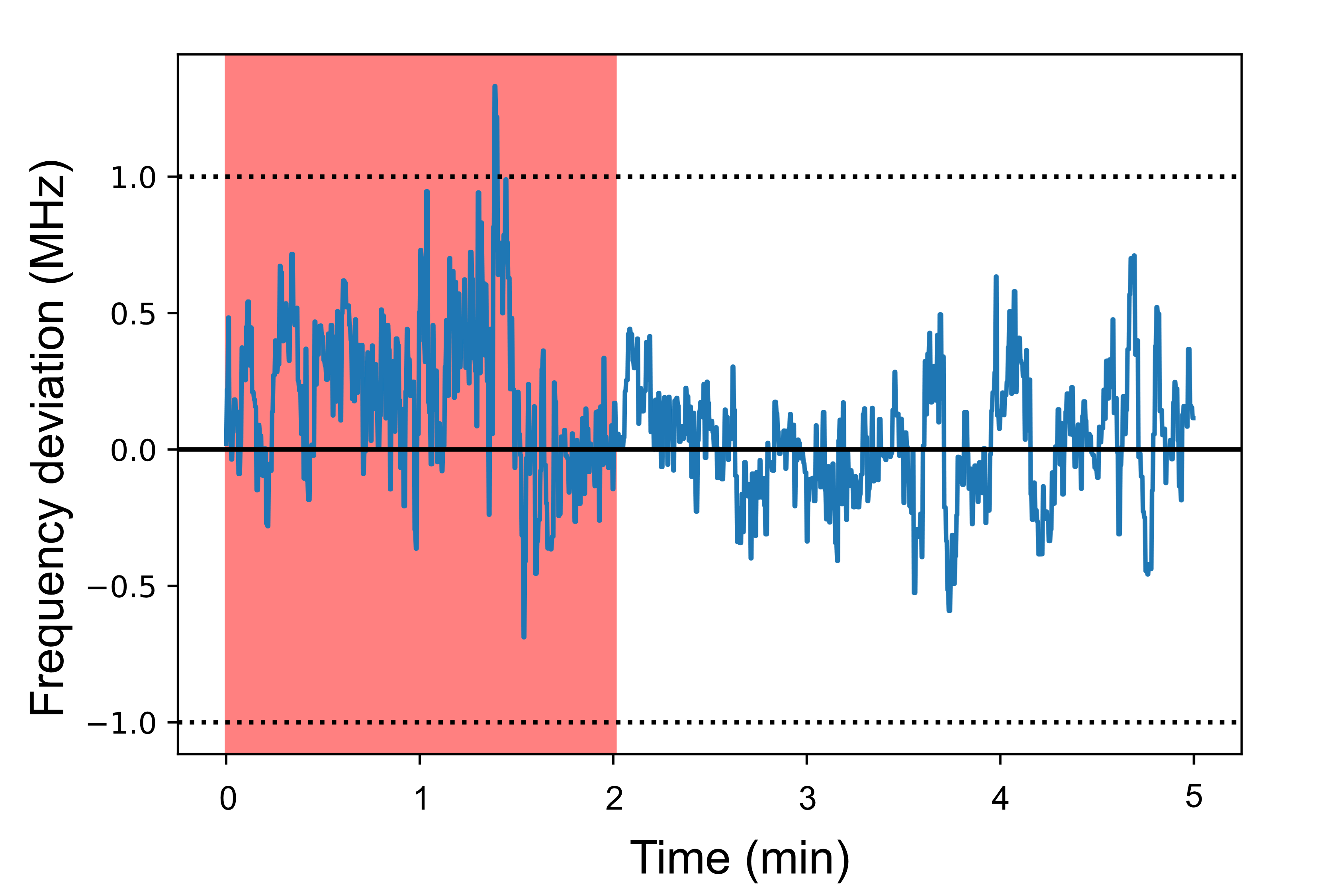}
    \caption{Results from mechanical robustness (vibration) test over a frequency tracking time of 5 min. Mechanical shock was applied for 2 min (red shaded region). During vibration, a mechanical shock up to 4g was applied to the module while locked. The frequency at the zero point of frequency deviation is 384.227981 THz. The frequency fluctuation reaches up to 1 MHz}
    \label{fig:mechanical_shock}
\end{figure}

We also conducted a vibration test on the module to demonstrate its mechanical stability under external shocks. 
Figure \ref{fig:mechanical_shock} shows the laser frequency response during manual shaking of the module, measured with the WLM for 5 minutes.
The vibration was applied by hand, and an accelerometer attached to the module recorded a peak stress of up to 4g. 
During the vibration period, the laser frequency deviated from the lock point by up to 1 MHz, and it immediately returned to the nominal lock point once the disturbance ceased.

\section{Conclusion}

In this work, we successfully developed an OFR module stabilized to a rubidium atomic transition. 
The module conforms to the standard 19‑inch rack format and demonstrates long‑term frequency stability close to the measurement limit, as well as robustness against external vibrations up to 4 g. 
It supports fully remote operation and can maintain stable performance for several months without user intervention, making it suitable for deployment in various experimental environments beyond conventional laboratory settings.

The CAD‑based design methodology and accompanying metadata enable straightforward distribution and reproduction of the module, allowing researchers to replicate or adapt the module with minimal effort. 
This redistributable design framework may support community sharing of optical blueprints and aid future efforts toward a modular design library for AMO systems.
Such an approach could facilitate rapid prototyping, collaborative refinement, and broader standardization of optical subsystems in future quantum and precision‑measurement experiments.

\begin{backmatter}

\bmsection{Funding}

This work was supported by the National Research Foundation of Korea (NRF) grant funded by the Korea government(MSIT) (2021R1F1A1062459, RS-2023-00302576, RS-2024-00466865, RS-2023-NR068116) and Institute of Information \& Communications Technology Planning \& Evaluation (ITTP) grant funded by the Korean government(MSIT) (RS-2022-II221040).

\bmsection{Disclosures}
The authors declare no conflicts of interest.

\bmsection{Data availability} 
The module documentation and design files are available on Github(\href{https://github.com/queti-at-skku/ofr-module}{https://github.com/queti-at-skku/ofr-module}), including a link to the associated Onshape CAD project.

\end{backmatter}


\begin{thebibliography}{10}
\newcommand{\enquote}[1]{``#1''}

\bibitem{ye_sub-doppler_nodate}
J.~Ye, L.~S. Ma, J.~L. Hall, \enquote{Sub-{Doppler} optical frequency reference at 1.064 mm by means of ultrasensitive cavity-enhanced frequency modulation spectroscopy of a {C2HD} overtone transition,} {\protect\JournalTitle{Optics Letters}} \textbf{21}, 1000-1002 (1996).

\bibitem{holzwarth_optical_2000}
R.~Holzwarth, T.~Udem, T.~W. Hänsch, \emph{et~al.}, \enquote{Optical {Frequency} {Synthesizer} for {Precision} {Spectroscopy},} {\protect\JournalTitle{Physical Review Letters}} \textbf{85}, 2264--2267 (2000).

\bibitem{hollberg_optical_2001}
L.~Hollberg, C.~Oates, E.~Curtis, \emph{et~al.}, \enquote{Optical frequency standards and measurements,} {\protect\JournalTitle{IEEE Journal of Quantum Electronics}} \textbf{37}, 1502--1513 (2001).

\bibitem{bartels_femtosecond-laser-based_2005}
A.~Bartels, S.~A. Diddams, C.~W. Oates, \emph{et~al.}, \enquote{Femtosecond-laser-based synthesis of ultrastable microwave signals from optical frequency references,} {\protect\JournalTitle{Optics Letters}} \textbf{30}, 667-669 (2005).

\bibitem{gill_optical_2005}
P.~Gill, \enquote{Optical frequency standards,} {\protect\JournalTitle{Metrologia}} \textbf{42}, S125--S137 (2005).

\bibitem{hall_nobel_2006}
J.~L. Hall, \enquote{Nobel {Lecture}: {Defining} and measuring optical frequencies,} {\protect\JournalTitle{Reviews of Modern Physics}} \textbf{78}, 1279--1295 (2006).

\bibitem{leibfried_quantum_2003}
D.~Leibfried, R.~Blatt, C.~Monroe, and D.~Wineland, \enquote{Quantum dynamics of single trapped ions,} {\protect\JournalTitle{Reviews of Modern Physics}} \textbf{75}, 281--324 (2003). 

\bibitem{bruzewicz_trapped-ion_2019}
C.~D. Bruzewicz, J.~Chiaverini, R.~McConnell, J.~M. Sage, \enquote{Trapped-{Ion} {Quantum} {Computing}: {Progress} and {Challenges},} {\protect\JournalTitle{Applied Physics Reviews}} \textbf{6}, (2019). 

\bibitem{henriet_quantum_2020}
L.~Henriet, L.~Beguin, A.~Signoles, \emph{et~al.}, \enquote{Quantum computing with neutral atoms,} {\protect\JournalTitle{Quantum}} \textbf{4}, 327 (2020).

\bibitem{wintersperger_neutral_2023}
K.~Wintersperger, F.~Dommert, T.~Ehmer, \emph{et~al.}, \enquote{Neutral atom quantum computing hardware: performance and end-user perspective,} {\protect\JournalTitle{EPJ Quantum Technology}} \textbf{10}, 32 (2023).

\bibitem{couteau_applications_2023}
C.~Couteau, S.~Barz, T.~Durt, \emph{et~al.}, \enquote{Applications of single photons to quantum communication and computing,} {\protect\JournalTitle{Nature Reviews Physics}} \textbf{5}, 326--338 (2023).

\bibitem{cozzolino_high-dimensional_2019}
D.~Cozzolino, B.~Da~Lio, D.~Bacco, L.~K. Oxenløwe, \enquote{High-{Dimensional} {Quantum} {Communication}: {Benefits}, {Progress}, and {Future} {Challenges},} {\protect\JournalTitle{Advanced Quantum Technologies}} \textbf{2}, 1900038 (2019). 

\bibitem{pezze_quantum_2018}
L.~Pezzè, A.~Smerzi, M.~K. Oberthaler, \emph{et~al.}, \enquote{Quantum metrology with nonclassical states of atomic ensembles,} {\protect\JournalTitle{Reviews of Modern Physics}} \textbf{90}, 035005 (2018).

\bibitem{giovannetti_advances_2011}
V.~Giovannetti, S.~Lloyd, L.~Maccone, \enquote{Advances in quantum metrology,} {\protect\JournalTitle{Nature Photonics}} \textbf{5}, 222--229 (2011).

\bibitem{zhang_microrod_2019}
W.~Zhang, F.~Baynes, S.~A. Diddams, S.~B. Papp, \enquote{Microrod {Optical} {Frequency} {Reference} in the {Ambient} {Environment},} {\protect\JournalTitle{Physical Review Applied}} \textbf{12}, 024010 (2019).

\bibitem{davila-rodriguez_compact_2017}
J.~Davila-Rodriguez, F.~N. Baynes, A.~D. Ludlow, \emph{et~al.}, \enquote{Compact, thermal-noise-limited reference cavity for ultra-low-noise microwave generation,} {\protect\JournalTitle{Optics Letters}} \textbf{42}, 1277-1280 (2017).

\bibitem{schuldt_high-performance_2016}
T.~Schuldt, K.~Döringshoff, A.~Milke, \emph{et~al.}, \enquote{High-{Performance} {Optical} {Frequency} {References} for {Space},} {\protect\JournalTitle{Journal of Physics: Conference Series}} \textbf{723}, 012047 (2016).

\bibitem{schkolnik_jokarus_2017}
V.~Schkolnik, K.~Döringshoff, F.~B. Gutsch, \emph{et~al.}, \enquote{{JOKARUS} - design of a compact optical iodine frequency reference for a sounding rocket mission,} {\protect\JournalTitle{EPJ Quantum Technology}} \textbf{4}, 1--10 (2017). 

\bibitem{strangfeld_prototype_2021}
A.~Strangfeld, S.~Kanthak, M.~Schiemangk, \emph{et~al.}, \enquote{Prototype of a compact rubidium-based optical frequency reference for operation on nanosatellites,} {\protect\JournalTitle{JOSA B}} \textbf{38}, 1885--1891 (2021). 

\bibitem{dinkelaker_autonomous_2017}
A.~N. Dinkelaker, M.~Schiemangk, V.~Schkolnik, \emph{et~al.}, \enquote{Autonomous frequency stabilization of two extended-cavity diode lasers at the potassium wavelength on a sounding rocket,} {\protect\JournalTitle{Applied Optics}} \textbf{56}, 1388--1396 (2017). 

\bibitem{pogorelov_compact_2021}
I.~Pogorelov, T.~Feldker, C.~D. Marciniak, \emph{et~al.}, \enquote{Compact {Ion}-{Trap} {Quantum} {Computing} {Demonstrator},} {\protect\JournalTitle{PRX Quantum}} \textbf{2}, 020343 (2021). 

\bibitem{zhang_compact_2018}
X.~Zhang, J.~Zhong, B.~Tang, \emph{et~al.}, \enquote{Compact portable laser system for mobile cold atom gravimeters,} {\protect\JournalTitle{Applied Optics}} \textbf{57}, 6545--6551 (2018). 

\bibitem{burke_compact_2005}
J.~H.~T. Burke, O.~Garcia, K.~J. Hughes, \emph{et~al.}, \enquote{Compact implementation of a scanning transfer cavity lock,} {\protect\JournalTitle{Review of Scientific Instruments}} \textbf{76}, (2005).

\bibitem{strangfeld_compact_2022}
A.~Strangfeld, B.~Wiegand, J.~Kluge, \emph{et~al.}, \enquote{Compact plug and play optical frequency reference device based on {Doppler}-free spectroscopy of rubidium vapor,} {\protect\JournalTitle{Optics Express}} \textbf{30}, 12039--12047 (2022). 

\bibitem{pahl_compact_2019}
J.~Pahl, A.~N. Dinkelaker, C.~Grzeschik, \emph{et~al.}, \enquote{Compact and robust diode laser system technology for dual-species ultracold atom experiments with rubidium and potassium in microgravity,} {\protect\JournalTitle{Applied Optics}} \textbf{58}, 5456--5464 (2019). 

\bibitem{chen_compact_2014}
Q.-F. Chen, A.~Nevsky, M.~Cardace, \emph{et~al.}, \enquote{A compact, robust, and transportable ultra-stable laser with a fractional frequency instability of $10^{-15}$,} {\protect\JournalTitle{Review of Scientific Instruments}} \textbf{85}, 113107 (2014).

\bibitem{myers_qubit_2025}
J.~Myers, C.~Caron, N.~Helaly, \emph{et~al.}, \enquote{Qubit operations using a modular optical system engineered with {PyOpticL}: a code-to-{CAD} optical layout tool,}  arXiv preprint arXiv:2501.14957 (2025). 

\bibitem{terra_ultra-stable_2016}
O.~Terra and H.~Hussein, \enquote{An ultra-stable optical frequency standard for telecommunication purposes based upon the {5S1}/2 → {5D5}/2 two-photon transition in rubidium,} {\protect\JournalTitle{Applied Physics B}} \textbf{122}, 27 (2016).

\bibitem{ye_hyperfine_1996}
J.~Ye, S.~Swartz, P.~Jungner, J.~L. Hall, \enquote{Hyperfine structure and absolute frequency of the $^{\textrm{87}}${Rb} {5P}$_{\textrm{3/2}}$ state,} {\protect\JournalTitle{Optics Letters}} \textbf{21}, 1280--1282 (1996). 

\bibitem{imanishi_frequency_2005}
S.~Imanishi, U.~Tanaka, S.~Urabe, \enquote{Frequency {Stabilization} of {Diode} {Laser} {Using} {Dichroic}-{Atomic}-{Vapor} {Laser} {Lock} {Signals} and {Thin} {Rb} {Vapor} {Cell},} {\protect\JournalTitle{Japanese Journal of Applied Physics}} \textbf{44}, 6767 (2005). 

\bibitem{noauthor_queti-at-skkuofr-module_nodate}
\enquote{queti-at-skku/ofr-module: {Optical} frequency module based on {Rb} {D2} transition,} \href{https://github.com/queti-at-skku/ofr-module}{https://github.com/queti-at-skku/ofr-module}  .

\bibitem{preston_dopplerfree_1996}
D.~W. Preston, \enquote{Doppler‐free saturated absorption: {Laser} spectroscopy,} {\protect\JournalTitle{American Journal of Physics}} \textbf{64}, 1432--1436 (1996).

\bibitem{business_onshape_nodate}
\enquote{Onshape, PTC} \href{https://www.onshape.com/en/}{https://www.onshape.com/en/}.

\bibitem{wiegand_linien_2022}
B.~Wiegand, B.~Leykauf, R.~Jördens, M.~Krutzik, \enquote{Linien: {A} versatile, user-friendly, open-source {FPGA}-based tool for frequency stabilization and spectroscopy parameter optimization,} {\protect\JournalTitle{Review of Scientific Instruments}} \textbf{93}, 063001 (2022).

\bibitem{bjorklund_frequency-modulation_1980}
G.~C. Bjorklund, \enquote{Frequency-modulation spectroscopy: a new method for measuring weak absorptions and dispersions,} {\protect\JournalTitle{Optics Letters}} \textbf{5}, 15--17 (1980). 

\bibitem{nakanishi_frequency-modulation_1987}
S.~Nakanishi, H.~Ariki, H.~Itoh, K.~Kondo, \enquote{Frequency-modulation spectroscopy of rubidium atoms with an {AlGaAs} diode laser,} {\protect\JournalTitle{Optics Letters}} \textbf{12}, 864--866 (1987). 

\bibitem{zi_laser_2017}
F.~Zi, X.~Wu, W.~Zhong, \emph{et~al.}, \enquote{Laser frequency stabilization by combining modulation transfer and frequency modulation spectroscopy,} {\protect\JournalTitle{Applied Optics}} \textbf{56}, 2649--2652 (2017). 

\bibitem{corwin_frequency-stabilized_1998}
K.~L. Corwin, Z.-T. Lu, C.~F. Hand, \emph{et~al.}, \enquote{Frequency-stabilized diode laser with the {Zeeman} shift in an atomic vapor,} {\protect\JournalTitle{Applied Optics}} \textbf{37}, 3295--3298 (1998). 

\bibitem{allan_statistics_1966}
D.~Allan, \enquote{Statistics of atomic frequency standards,} {\protect\JournalTitle{Proceedings of the IEEE}} \textbf{54}, 221-230 (2005).

\bibitem{chen_stable_2022}
T.~Chen, J.~Kim, M.~Kuzyk, \emph{et~al.}, \enquote{Stable {Turnkey} {Laser} {System} for a {Yb}/{Ba} {Trapped}-{Ion} {Quantum} {Computer},} {\protect\JournalTitle{IEEE Transactions on Quantum Engineering}} \textbf{3}, 1--8 (2022).

\end{thebibliography}
\end{document}